\documentclass[twocolumn,showpacs,preprintnumbers,pre,floatfix]{revtex4}

\newcommand{\BE}{\begin{equation}}
\newcommand{\EE}{\end{equation}}
\newcommand{\BA}{\begin{eqnarray}}
\newcommand{\EA}{\end{eqnarray}}

\newcommand{\BW}{\begin{widetext}}
\newcommand{\EW}{\end{widetext}}

\input epsf.tex

\usepackage{graphicx}
\usepackage{dcolumn}
\usepackage{bm}

\begin{document}

\title{Group velocity of the acoustic eigen-modes in sonic crystals}

\author{Zhen Ye}\email{zhen@shaw.ca}
\affiliation{Wave Phenomena Laboratory, Department of Physics,
National Central University, Chungli, Taiwan 32054}
\date{\today}

\begin{abstract}

In this brief report, the group velocity of the eigen-modes in
sonic crystals is derived, and shown to equal the averaged energy
velocity of the eigen-modes. How the group velocity can be used to
describe acoustic energy flows in sonic crystals is discussed.

\end{abstract}

 \maketitle

The phenomenon of band structures, revealed as waves propagate
through periodic structures, was put on a solid foundation in the
context of Bloch's theorem\cite{Bloch}. It was first studied for
electronic systems. Since the central physics behind electronic
band structures lies in the wave nature of electrons, it is
natural to extend the study to any wave systems in general, and to
electromagnetic (EM) and acoustic systems in particular.

Not excluding other efforts, band structures were earlier
addressed for general waves in periodic structures by
Brillouin\cite{Bri}, then to EM waves by Yariv and
Yeh\cite{Yariv}, later by Yablonovitch\cite{Yab} and
John\cite{John}. Since then, the study of EM waves in periodic
structures has been booming, eventually extended to the
exploration of acoustic waves in periodic structures (e.~g.
Refs.~\cite{Dowling,sanchez1,sanchez2,Kush}), leading to the
establishment of the field of photonic crystals (PCs) and sonic
crystals (SCs).

Sonic crystals (SCs) are made of periodically structured materials
which are sensitive to acoustic waves, and have been studied both
intensively and extensively. The exciting phenomenon of band
structures in sonic crystals allows for many possible
applications. It has been recognized that SCs could be used as
sound shields, acoustic filters, acoustic flow guiding, sonic
crystal lenses, and so
on\cite{sanchez1,sanchez2,caba1,rober,sanchis,kuswa,Ye_lens,Ye_APL}.

One of the top issues in the research of acoustic crystals is how
to describe the propagation of acoustic waves in crystal
structures, by analogy with the photonic crystals. The common
theoretical approach to electromagnetic propagation in periodic
media has been given in Ref.~\cite{Yariv}, and may be summarized
as follows\cite{Ye_Energy}. The Maxwell equations are first
derived for waves in periodic media. By Bloch theorem, the
solution can be expanded in terms of Bloch waves. The solution is
then substituted into the governing equations to obtain an
eigen-equation that determines the dispersion relations between
the frequency and the wave vector that lies within the first
Brillouin zone for the eigen-modes. These relations are termed as
frequency band structures. Since it has been
proved\cite{Yariv,Sakoda} that the averaged energy velocity
$\vec{v}_e$ equals the group velocity which can be obtained as the
gradient of the dispersion relations with the respect to the space
of wave vectors, i.~e. $\vec{v}_g \equiv \nabla_{\vec{K}}\omega$,
the investigation of electromagnetic propagation in periodic
structures is thus reduced to the calculation of the group
velocity from the band structures.

Simply due to similarities between the electromagnetic and
acoustic waves, it is naturally expected that the above
prescription can be automatically adopted for sonic crystals, and
this recipe has been indeed used for SCs in the literature (e.~g.
Ref.~\cite{CTChan}). To the best of our knowledge, however, so far
there has been no attempt in the literature to prove this analogy
and connection between PCs and SCs. Here, we wish to bridge the
gap. In this paper, we will provide a proof that in analogy with
PCs, the acoustic energy velocity of the eigen-modes of a sonic
crystal averaged within a unit cell also equals the group velocity
of these eigen-modes, defined as $\vec{v}_g \equiv
\nabla_{\vec{K}}\omega$. Moreover, we will discuss how the group
velocity can be used to describe acoustic energy flows in sonic
crystals. In addition, following the work on photonic
crystals\cite{Luan}, it will be shown that measuring the unit-cell
volume averaged acoustic energy velocity can be reduced to the
measurement of the corresponding unit-cell surface averaged
components. Such a reduction is expected to facilitate the
practical measurement of acoustic energy flows.


{\it\bf General} - In the linear region, the propagation of
acoustic waves are governed by three physical principles: Newton's
second law, conservation law, and equation of states. And acoustic
waves are characterized by three major physical quantities:
variations in pressure and mass density, and the displacement
velocity, denoted respectively by $\delta p$, $\delta \rho$, and
$\vec{u}$. The governing equations are then \BE \rho\nabla
\cdot\vec{u} = - \frac{\partial \delta\rho}{\partial t}, \  \
\frac{\partial \vec{u}}{\partial t} = -\frac{1}{\rho}\nabla
(\delta p),\ \mbox{and}\ (\delta \rho) = \frac{1}{c^2}(\delta
p).\EE These equations lead to \BE \nabla \cdot\vec{u} =
-\frac{1}{c^2\rho}\frac{\partial \delta p}{\partial t} \EE So we
have two main equations \BE \nabla \cdot\vec{u} =
-\frac{1}{c^2\rho}\frac{\partial p}{\partial t} \label{eq:scm1}\EE
and \BE \frac{\partial {\vec{u}}}{\partial t} =
-\frac{1}{\rho}\nabla p,\label{eq:scm2}\EE where we ignore
`$\delta$'.

From Eqs.~(\ref{eq:scm1}) and (\ref{eq:scm2}), we have the wave
equation \BE \nabla\cdot\left(\frac{1}{\rho}\nabla p\right) -
\frac{1}{c^2\rho}\frac{\partial^2 p}{\partial^2 t} = 0.\EE

Multiplying Eq.~(\ref{eq:scm1}) with $p^\star$, and multiplying
Eq.~(\ref{eq:scm2}) with $\vec{u}^\star$, and then adding the
correspondingly the complex conjugated equations, we will get \BE
\frac{1}{2}\nabla\cdot( \vec{u}p^\star + \vec{u}^\star p) =
-\frac{1}{2c^2\rho}\frac{\partial |p|^2}{\partial t}
-\frac{\rho}{2} \frac{\partial |\vec{u}|^2}{\partial t} \EE This
equation can be written as \BE \nabla \cdot \vec{J} =
-\frac{\partial}{\partial t} U, \label{eq:cons}\EE with $ \vec{J}
\equiv \mbox{Re}[p^\star \vec{u}],$ and $ U \equiv
\frac{1}{2c^2\rho} |p|^2 + \frac{\rho}{2}|\vec{u}|^2$.
Eq.~(\ref{eq:cons}) represents the energy conservation. The
quantities $\vec{J}$ and $U$ are the acoustic intensity and energy
respectively.


{\it \bf Proof of $\nabla_{\vec{K}}\omega = \vec{v}_e$} -  Now we
consider sonic crystals. We will consider three dimensional cases;
the extension to two dimensions is straightforward. The band
structures can be readily computed by the plane-wave method. Here
we assume that the band structure is known. Readers may refer to
Ref.~\cite{Kush} for details. We will prove that the acoustic
energy velocity of the eigen-modes of a sonic crystal averaged
within a unit cell equals the group velocity of these eigen-modes.

Considering monochromatic waves in sonic crystals, from
Eqs.~(\ref{eq:scm1}) and (\ref{eq:scm2}) \BA
\nabla \cdot\vec{u} &=& \frac{i\omega}{c^2\rho}p,\label{eq:scc3}\\
\nabla p &=& i\omega\rho\vec{u}\label{eq:scc4}.\EA Hereafter we
drop out the time factor $e^{-i\omega t}$.

According to Bloch's theorem, waves of the eigen-modes in periodic
structures can be expressed in terms of Bloch functions, \BE
\vec{u} = \vec{u}_{\vec{K}} e^{i\vec{K}\cdot\vec{r}},\ \ \ p =
p_{\vec{K}}e^{i\vec{K}\cdot\vec{r}}.\EE Taking these two equations
into Eqs.~(\ref{eq:scc3}) and (\ref{eq:scc4}), we have \BA \nabla
\cdot\vec{u}_{\vec{K}} + i\vec{K}\cdot \vec{u}_{\vec{K}} &=&
\frac{i\omega}{c^2\rho}p_{\vec{K}}, \label{eq:scc5}\\
\nabla p_{\vec{K}} + i\vec{K} p_{\vec{K}} &=&
i\omega\rho\vec{u}_{\vec{K}}.\label{eq:scc6}\EA

Considering an arbitrary infinitesimal change in $\vec{K}$, which
induces the changes in $\omega$, $u$, and $p$,
Eqs.~(\ref{eq:scc5}) and (\ref{eq:scc6}) become \begin{widetext}
\BA \nabla \cdot\delta\vec{u}_{\vec{K}} + i\delta\vec{K}\cdot
\vec{u}_{\vec{K}} + i\vec{K}\cdot \delta\vec{u}_{\vec{K}}&=&
\frac{i}{c^2\rho}\delta\omega p_{\vec{K}}+ \frac{i}{c^2\rho}\omega\delta p_{\vec{K}}, \label{eq:scc7}\\
\nabla \delta p_{\vec{K}} + i\delta\vec{K}
p_{\vec{K}}+i\vec{K}\delta p_{\vec{K}} &=&
i\rho\delta\omega\vec{u}_{\vec{K}} +
i\rho\omega\delta\vec{u}_{\vec{K}}.\label{eq:scc8}\EA\end{widetext}

The corresponding complex conjugates are \BW\BA \nabla
\cdot\delta\vec{u}^\star_{\vec{K}} - i\delta\vec{K}\cdot
\vec{u}^\star_{\vec{K}} - i\vec{K}\cdot
\delta\vec{u}^\star_{\vec{K}}&=&
-\frac{i}{c^2\rho}\delta\omega p^\star_{\vec{K}} - \frac{i}{c^2\rho}\omega\delta p^\star_{\vec{K}}, \label{eq:scc9}\\
\nabla \delta p^\star_{\vec{K}} - i\delta\vec{K} p^\star_{\vec{K}}
- i\vec{K}\delta p^\star_{\vec{K}} &=& -
i\rho\delta\omega\vec{u}^\star_{\vec{K}} -
i\rho\omega\delta\vec{u}^\star_{\vec{K}}.\label{eq:scc10}\EA \EW

Multiplying Eq.~(\ref{eq:scc7}) with $p^\star_{\vec{K}}$, and
Eq.~(\ref{eq:scc9}) with $p_{\vec{K}}$, then making a subtraction
between the two resulting equations, we have \BW\BE 2i
\mbox{Im}[p^\star_{\vec{K}}\nabla\cdot\delta\vec{u}_{\vec{K}}] +
2i\delta\vec{K}\cdot\vec{J}_{\vec{K}} +
2i\vec{K}\cdot\mbox{Re}[p^\star_{\vec{K}}\delta\vec{u}_{\vec{K}}]
= \frac{2i}{c^2\rho}\delta\omega |p_{\vec{K}}|^2 +
\frac{2i}{c^2\rho}\omega\mbox{Re}[p_{\vec{K}}\delta
p^\star_{\vec{K}}]\label{eq:scc11},\EE \EW where
$$\vec{J}_{\vec{K}} =
\mbox{Re}[p^\star_{\vec{K}}\vec{u}_{\vec{K}}].$$

Multiplying Eq.~(\ref{eq:scc8}) with $\vec{u}^\star_{\vec{K}}$,
and Eq.~(\ref{eq:scc10}) with $\vec{u}_{\vec{K}}$, then making a
subtraction between the two resulting equations, we have \BW\BE
2i\mbox{Im}[\vec{u}^\star_{\vec{K}}\cdot \nabla \delta
p_{\vec{K}}] +  2i\delta\vec{K}\cdot\vec{J}_{\vec{K}} +
2i\vec{K}\cdot\mbox{Re}[\vec{u}^\star_{\vec{K}} \delta
p_{\vec{K}}] = 2i\rho \delta\omega |u_{\vec{K}}|^2 + 2i \rho\omega
\mbox{Re}[\vec{u}^\star_{\vec{K}}\cdot \delta
\vec{u}_{\vec{K}}].\label{eq:scc12}\EE\EW

Adding Eqs.~(\ref{eq:scc11}) and (\ref{eq:scc12}), and taking into
account Eqs.~(\ref{eq:scc5}) and (\ref{eq:scc6}), we have \BW\BE
4i\delta\vec{K}\cdot\vec{J}_{\vec{K}} =
\frac{2i}{c^2\rho}\delta\omega |p_{\vec{K}}|^2 +  2i\rho
\delta\omega |u_{\vec{K}}|^2 + 2i\mbox{Im}[\nabla\cdot(
p^\star_{\vec{K}}\delta\vec{u}_{\vec{K}}) +\nabla\cdot(
\vec{u}^\star_{\vec{K}}\delta p_{\vec{K}})]. \label{eq:scf1}\EE\EW
Due to the periodicity, the unit cell integration for any periodic
function $\vec{A}$ can be shown to be zero, i.~e. $ \int_C
d\vec{r} \nabla\cdot\vec{A} = 0.$ Now performing the unit cell
integration for Eq.~(\ref{eq:scf1}), we have \BE
\left\langle\frac{1}{2c^2\rho} |p_{\vec{K}}|^2 +  \frac{\rho}{2}
|u_{\vec{K}}|^2\right\rangle \delta\omega
=\delta\vec{K}\cdot\langle\vec{J}_{\vec{K}}\rangle,
\label{eq:scf2}\EE where $\langle A\rangle \equiv
\frac{1}{V_c}\int_C d\vec{r} A$, with $V_c$ being the volume of
the unit cell.

We define the unit-cell volume averaged energy velocity as \BE
\vec{v}_{e,\vec{K}} \equiv
\frac{\langle\vec{J}_{\vec{K}}\rangle}{\left\langle\frac{1}{2c^2\rho}
|p_{\vec{K}}|^2 +  \frac{\rho}{2}
|u_{\vec{K}}|^2\right\rangle}.\EE Then we have \BE \delta\omega =
\vec{v}_{e,\vec{K}}\cdot\delta\vec{K}.\EE

Since \BE \delta\omega =
\nabla_{\vec{K}}\omega\cdot\delta\vec{K},\EE we have finally
proved the identity for the group velocity defined as $\vec{v}_{g,
\vec{K}} \equiv \nabla_{\vec{K}}\omega$: \BE \vec{v}_{g, \vec{K}}
= \vec{v}_{e,\vec{K}} = \nabla_{\vec{K}}\omega .\EE When the index
$\vec{K}$ for the eigen-mode is ignored, we simply have \BE
\vec{v}_g = \vec{v}_e,\EE with $\vec{v}_g =
\nabla_{\vec{K}}\omega,$ which was first derived for PCs.

From the above, we know that $\nabla_{\vec{K}}\omega$ may describe
the averaged energy flow in sonic crystals. For example, the
direction of averaged energy follows that of
$\nabla_{\vec{K}}\omega$. As such, in the study of photonic
crystals, $\nabla_{\vec{K}}\omega$ is regarded as the key quantity
in discerning the photonic flows. Here, however, we must stress
that the average of the energy velocity is performed over the
whole unit cell of periodic media, under what condition such
averaged energy flow can depict the actual energy flow in periodic
media is worth considering. Moreover, since in actual measurements
it is often hard to detect the physical quantities inside the
periodic media, how to obtain the group velocity without having to
put a detector into the media needs also to be considered. In the
following, we will consider these two questions.

First, due to the periodicity, the local acoustic current of the
eigen-modes can be written as \BE \vec{J}_{\vec{K}} =
\sum_{\vec{G}} \vec{J}_{\vec{K}}(\vec{G})
e^{i\vec{G}\cdot\vec{r}},\EE where $\vec{G}$ denotes the
reciprocal vectors. It is shown that \BA \frac{1}{V_c}\int_C
d\vec{r} \vec{J}_{\vec{K}} &=& \frac{1}{V_c}\int_C d\vec{r}
\sum_{\vec{G}}\vec{J}_{\vec{K}}(\vec{G})e^{i\vec{G}\cdot\vec{r}}
\nonumber \\ &=& \sum_{\vec{G}}\vec{J}_{\vec{K}}(\vec{G})
\delta_{0,\vec{G}} = \vec{J}_{\vec{K}}(0). \label{eq:r33}\EA From
this result, we clearly see that when and only when the component
with $\vec{G}=0$ dominates, the group velocity $\vec{v}_g =
\nabla_{\vec{K}}\omega$ can represent well the actual wave flows.
In this case, the eigen-modes will behave effectively as a plane
wave in a free space. Some recent studies indicate that this may
not be generally valid, and have caused in certain circumstances
some confusions about whether waves are diffracted or refracted,
referring to the discussions in Refs.~\cite{PLA,CJP}.
Additionally, when a crystal is ensonified by an external wave,
many eigen-modes may be excited for a given frequency, adding
complications to the determination of the actual acoustic
propagation.


{\it \bf Reduction of integration} - Second, it can be shown that
the unit volume averaged acoustic intensity $ \langle
\vec{J}_{\vec{K}}\rangle = \frac{1}{V_c}\int_C d\vec{r}
\vec{J}_{\vec{K}}$ can be represented by surface averaged acoustic
intensity. This is important, since it will allow us to measure
certain physical quantities at surfaces rather than putting
detectors into the volume of samples. To briefly show this, we
will take the procedure stemming from the excellent
note~\cite{Luan}. For details, readers should refer to
Ref.~\cite{Luan}.

We consider an acoustic lattice with the base lattice vectors
$\vec{a}_1$, $\vec{a}_2$, and $\vec{a}_3$. The base vectors for
the reciprocal lattice $\vec{b}_1, \vec{b}_2$ and $\vec{b}_3$ are
determined by $\vec{a}_i\cdot\vec{b}_j = 2\pi\delta_{ij}$. We will
show that \BE \langle \vec{J}_{\vec{K}}\rangle\cdot \hat{b}_i =
\frac{1}{S_i}\int_{S_i}ds
\vec{J}_{\vec{K}}\cdot\hat{b}_i,\label{eq:r1}\EE where $\hat{b}_i$
is the unit vector of $\vec{b}_i$, given by $\hat{b}_i =
\frac{V_c}{2\pi S_i} \vec{b}_i$, and $S_i$ denotes the three
surfaces of a unit cell, characterized by
$\vec{a}_2\times\vec{a}_3$, $\vec{a}_3\times\vec{a}_1$,
$\vec{a}_1\times\vec{a}_2$ respectively.

For the stationary case, i.~e. $ \nabla\cdot\vec{J}_{\vec{K}} =
0$, we have from Eq.~(\ref{eq:r33}) \BE
\vec{G}\cdot\vec{J}_{\vec{K}}(\vec{G})=0\label{eq:GJ}\EE

To show the identity in Eq.~(\ref{eq:r1}), we consider the surface
$S_3$ as the example. Writing $\vec{G} = \sum_{i=1}^3
m_{i}\vec{b}_i$, we have \BA \frac{1}{S_3} \int_{S_3} ds
\vec{J}_{\vec{K}}\cdot\hat{b}_3 &=&
\frac{1}{S_3}\sum_{\vec{G}}\vec{J}_{\vec{K}}(\vec{G})\cdot\hat{b}_3
\int_{S_3}ds e^{i\vec{G}\cdot\vec{r}} \nonumber\\
&=&
\frac{1}{S_3}\sum_{\vec{G}}\vec{J}_{\vec{K}}(\vec{G})\cdot\hat{b}_3
S_3\delta_{0, m_1}\delta_{0,m_2}\nonumber \\ &=&
\sum_{m_3}\vec{J}_{\vec{K}}(m_3\vec{b}_3)\cdot\hat{b}_3.\label{eq:r30}\EA
Applying Eq.~(\ref{eq:GJ}) to Eq.~(\ref{eq:r30}), we arrive at \BE
\frac{1}{S_3} \int_{S_3} ds \vec{J}_{\vec{K}}\cdot\hat{b}_3 =
\vec{J}_{\vec{K}}(0)\cdot\hat{b}_3.\label{eq:r2}\EE

From Eq.~(\ref{eq:r33}) and by comparing to Eq.~(\ref{eq:r2}), we
get \BE \langle \vec{J}_{\vec{K}}\rangle\cdot \hat{b}_3 =
\frac{1}{S_i}\int_{S_i}ds \vec{J}_{\vec{K}}\cdot\hat{b}_3.\EE
Following the same procedure, we can have the similar results for
the other two surfaces. Thereby we prove \BE \langle
\vec{J}_{\vec{K}}\rangle\cdot \hat{b}_i =
\frac{1}{S_i}\int_{S_i}ds
\vec{J}_{\vec{K}}\cdot\hat{b}_i.\label{eq:r4}\EE This result
indicates that the components of the volume averaged acoustic
intensity can be reduced to the surface averaged counterparts.
Clearly, this is expected to be very useful to actual
measurements.

In summary, we have provided a proof that in analogy with PCs, the
acoustic energy velocity of the eigen-modes of a sonic crystal
averaged within a unit cell also equals the group velocity of
these eigen-modes, defined as $\vec{v}_g \equiv
\nabla_{\vec{K}}\omega$. We have also discussed the conditions
that the group velocity can be used to describe acoustic energy
flows in sonic crystals. Then it is shown that the volume averaged
acoustic energy velocity can be reduced to the surface averaged
components.

{\bf Acknowledgements} Discussion with and help from P.-G. Luan
are greatly appreciated. This work received support from National
Science Council of Republic of China.

\end{document}